# Posttreatment Effects on the Crystal Structure and Superconductivity of Ca-free Double-Layered Cuprate $Sr_2SrCu_2O_{4+y}F_{2-y}$

Hiroki Ninomiya[1]*‡, Kenji Kawashima[1,2]‡, Hiroshi Fujihisa[1], Shigeyuki Ishida[1], Hiraku Ogino[1], Yoshiyuki Yoshida[1], Hiroshi Eisaki[1], Yoshito Gotoh[1], and Akira Iyo[1]

*[1]National Institute of Advanced Industrial Science and Technology (AIST), 1-1-1 Umezono, Tsukuba, Ibaraki 305-8568, Japan*

*[2]IMRA Japan Co., Ltd., 2-36 Hachiken-cho, Kariya, Aichi 448-8650, Japan*

ABSTRACT. We report the effects of low-temperature postannealing on the structural and superconducting properties of the recently discovered Ca-free double-layered cuprate, $Sr_2SrCu_2O_{4+y}F_{2-y}$. Although the as-synthesized sample prepared under high pressure has a tetragonal structure with a rock-salt-type blocking layer (the so-called *T*-phase), we found that the symmetry of the structure lowered to that of an orthorhombic system when annealed with $CuF_2$. The structural refinements reveal that such a topochemical reaction leads not only to the removal of excess $O^{2-}$ from the apical site but also to the intercalation of extra $F^{-}$ into the interstitial site. The orthorhombic phase exhibits bulk superconductivity at a critical temperature of 107 K, which is significantly higher than that of the *T*-phase (~50 K). Meanwhile, the *T*-phase turns into another structure possessing a fluorite-type blocking layer without apical fluorine (known as the *T*´-phase) by annealing without $CuF_2$. Density functional theory calculations show that the *T*´-phase is more stable than the *T*-phase. This is the first report on the formation of a *T*´-type double-layered cuprate. Furthermore, structural stability of the three phases of $Sr_2SrCu_2O_{4+y}F_{2-y}$ is discussed in terms of lattice matching between the blocking and conducting layers.

## 1. INTRODUCTION

Cuprate superconductors, whose structure contains two-dimensional $CuO_2$ layers, have maintained the highest superconducting critical temperature ($T_c$) under ambient pressures for more than 30 years since their discovery.[1,2] In particular, superconducting cuprates with $T_c$ above 100 K possess a multilayered structure, where each of the multiple $CuO_2$ layers is separated by a Ca ion.[3,4] Meanwhile, our recent study[5] has yielded a new class of cuprate superconductors with the chemical formulae $Sr_2SrCu_2O_{4+y}F_{2-y}$, (Hg, Re)$Sr_2SrCu_2O_y$, $TlSr_2SrCu_2O_y$, and (B, C)$Sr_2SrCu_2O_y$. Unlike conventional multilayered systems such as $Ba_2CaCu_2O_{4+y}F_{2-y}$ and $HgBa_2CaCu_2O_y$, a novel structural feature of the Ca-free cuprates is that the divalent $Sr^{2+}$ plays a role in separating the $CuO_2$ layers and in forming a blocking layer.

The Ca-free cuprate $Sr_2SrCu_2O_{4+y}F_{2-y}$ is a double-layered version of the known single-layered cuprate $Sr_2CuO_2F_{2+\delta}$ with $T_c$ = 46 K.[6] This phase is stabilized via the solid-state reaction of the nominal composition $Sr_2SrCu_2(O_{4\pm z+y}H_z)F_{2-y}$ ($z$ = 0.2, $y$ = 0.2–0.6) under high pressures (HP). (It is to be noted that a small amount of $Sr(OH)_2$ is added as a reaction accelerator.) The as-synthesized sample with the tetragonal space group *I*4/*mmm* (No. 139) possesses a rock-salt-type blocking layer (also known as the *T*-phase), which is identical to the typical $Ba_2CaCu_2O_{4+y}F_{2-y}$ oxyfluoride cuprate.[7,8] According to our previous study,[5] the crystal structure changes to that of an orthorhombic system by topochemical fluorination when the *T*-phase is annealed at 453 K with $CuF_2$. For the appearance of superconductivity, while the as-synthesized sample has a $T_c$ of 60 K, the fluorinated sample exhibits a $T_c$ of 107 K, which is significantly higher than that before fluorination and comparable to $T_c$ = 108 K of $Ba_2CaCu_2O_{4+y}F_{2-y}$.[5,7] Additionally, we have also reported that the as-synthesized sample with the *T*-phase is structurally unstable under ambient pressure and seems to change to a different structure.[5] As $Sr_2SrCu_2O_{4+y}F_{2-y}$ is a member of a new family of high-$T_c$ superconducting materials, it is important to unravel the various crystal structures appearing in it and its superconducting properties.

Herein, we elucidate the effects of two postannealing processes, namely, low-temperature annealing with and without a fluorinating agent, on $Sr_2SrCu_2O_{4+y}F_{2-y}$. For this purpose, we first examined the composition ($y$) dependence of the structural parameters and the critical temperature of the sample before annealing. Second, we investigated the evolution of the structure and superconductivity in the different postannealed samples and compared them with the results of the as-synthesized sample. Third, we conducted structural analyses on each sample using the Rietveld method based on density functional theory (DFT) structural optimization and molecular dynamics (MD) simulations. Finally, we discuss the structural stability of this material and compare its structural parameters with those of other single- and multilayered oxyfluoride cuprates.

## 2. EXPERIMENTAL AND COMPUTATIONAL DETAILS

### *2.1 Material Preparation and the Postannealing Process*

Polycrystalline $Sr_2SrCu_2O_{4+y}F_{2-y}$ samples were prepared using an HP synthesis technique. As described in Ref. 5, to obtain the target phase predominantly, we added hydrogen to the starting composition using $Sr(OH)_2$. The nominal composition, $Sr_2SrCu_2O_{4.6}F_{1.4}H_{0.2}$ ($y$ = 0.2, 0.4, and 0.6), was sintered at 1173 K for 1 h and at a pressure of 3.4 GPa. For the posttreatment, copper (II) difluoride ($CuF_2$) was used as the fluorinating agent. The as-synthesized samples were powdered and pressed into pellets in a $N_2$-filled glovebox. The pellet was sealed in a quartz tube with a $CuF_2$ pellet. The typical weight ratio of the sample to $CuF_2$ was 1:3. The ampoule was heated to 453 K and this temperature was maintained for 12 h. The annealing without $CuF_2$ was performed in an evacuated quartz tube at 533–548 K for approximately 2 days.

### *2.2. Structure Characterization and Refinements*

The obtained samples were characterized by powder X-ray diffraction (XRD) at a temperature of approximately 293 K with Cu-K$\alpha$ radiation using a diffractometer (Rigaku, Ultima IV) equipped with a high-speed detector system (Rigaku, D/teX Ultra). Intensity ($I$) data were collected over the 5–80° range in 0.02° steps. The lattice constants of the samples with each $y$ were calculated using the least-squares method. For the structure analysis data, we performed measurements over the 5–140° range in 0.01° steps. The crystallographic parameters of the fluorinated samples were refined by Rietveld analysis using BIOVIA Materials Studio (MS) Reflex software (version 2018 R2).[9] Because it is difficult to determine the exact positions of the lightweight O and F elements from XRD, optimal crystal structures were calculated via DFT using the MS-CASTEP code[10] with GGA-PBE exchange-correlation functionals[11] and ultrasoft pseudopotentials.[12] An energy cutoff of 571 eV was used for the plane wave basis set. Moreover, the Monkhorst–Pack grid separation[13] was set to approximately 0.04 $Å^{-1}$. For MD simulations, we used a Nosé–Hoover thermostat[14] and a Parrinello–Rahman barostat[15] in a NPT ensemble. A time step of 4 fs was employed, and the simulation time was 8 ps. For the sample annealed without $CuF_2$, the XRD data were refined using a $T'$-type structure model optimized by DFT calculation under the same conditions as mentioned earlier.

### *2.3. Superconducting Property Measurements*

The temperature ($T$) dependence of the magnetic susceptibilities ($\chi$) was measured using a SQUID magnetometer (Quantum Design, MPMS-XL) in an applied magnetic field ($H$) of 10 Oe. The sample was sealed in a plastic capsule, and these experiments were performed using both zero-field cooling (ZFC) and field cooling (FC) processes. Because the structure of the $T$-phase gradually changes with time, the XRD and $\chi$–$T$ measurements were carried out immediately after the synthesis.

## 3. RESULTS

### *3.1. Superconductivity of the As-Synthesized $Sr_2SrCu_2O_{4+y}F_{2-y}$*

**Figure 1(a),(b)** illustrates the variation in lattice constants ($a$ and $c$) and the $T_c$ of the as-synthesized samples with the nominal composition of $Sr_2SrCu_2O_{4+y}F_{2-y}$. For comparison, corresponding data for $Ba_2CaCu_2O_{4+y}F_{2-y}$[7] are also plotted. The (200) and (0012) peaks of the XRD profile and the zero-field-cooled susceptibility data are shown in Figure S1. Here, an increase in $y$ indicates that the formal valence of Cu increases with $Cu^{(2+(y/2))+}$. Note that there possibly exists a discrepancy between the nominal and actual compositions of O and F; however, it is difficult to quantify their content experimentally. For $Ba_2CaCu_2O_{4+y}F_{2-y}$, with increasing $y$, the length of the $a$-axis decreased by 0.15%, while that of the $c$-axis increased by 0.57%. Similarly, the $c$-axis length of $Sr_2SrCu_2O_{4+y}F_{2-y}$ also increased by 0.32%, while the $a$-axis length remained almost unchanged (or decreased slightly by 0.03%). These results suggest that the carrier concentration changed systematically when the initial $y$ value was varied. Considering that the $Sr^{2+}$ ion, which is larger than the $Ca^{2+}$ ion, separates the $CuO_2$ layer, $Sr_2SrCu_2O_{4+y}F_{2-y}$ has a possible limitation of in-plane shrinkage. As shown in Figure 1(b), the $T_c$ of $Ba_2CaCu_2O_{4+y}F_{2-y}$ exhibited the maximum value (108 K), indicating the formation of an optimally doped state at approximately $y = 0.4$.[7] Contrastingly, the $T_c$ of $Sr_2SrCu_2O_{4+y}F_{2-y}$ monotonically decreased up to $y = 0.6$. Therefore, we speculate that the sample with $y = 0.2$ is in the nearly optimal doped or slightly overdoped state, and a further increase in $y$ will shift the sample to a heavily overdoped region.

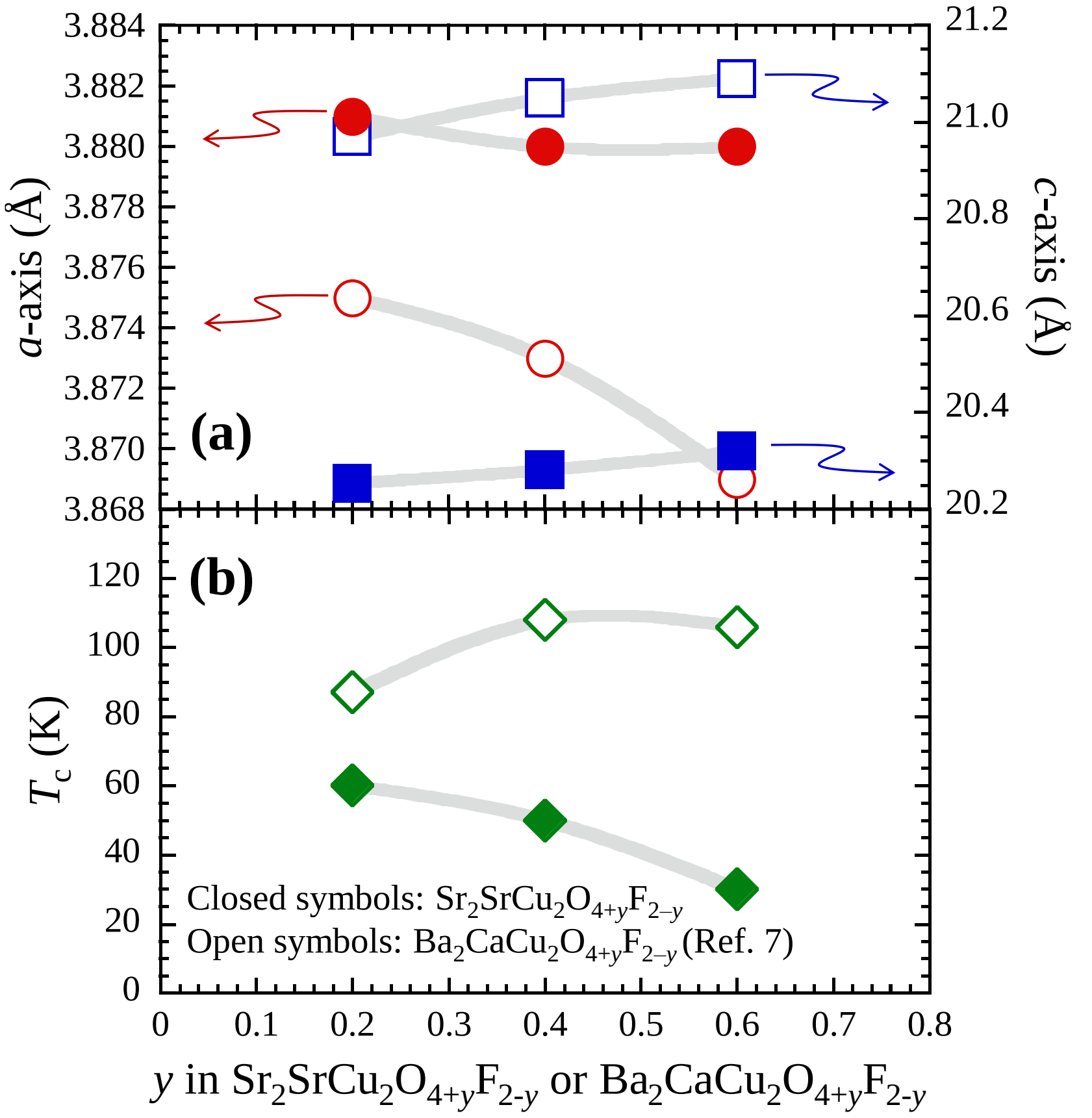

**Figure 1**: (a) Lattice constants and (b) $T_c$ as a function of the nominal $y$ in $Sr_2SrCu_2O_{4+y}F_{2-y}$ and $Ba_2CaCu_2O_{4+y}F_{2-y}$.[7]

### *3.2. Low-Temperature Postannealing Effects*

**Figure 2(a)–(c)** shows the XRD patterns of the as-synthesized sample and the annealed sample with and without $CuF_2$, respectively. The data in the upper panel were well indexed with a body-centered tetragonal structure with the *I*4/*mmm* space group (No. 139) referred to as the $Ca_2CaCu_2O_4Cl_2$-type model.[16] The other small peaks originate from impurities such as $Sr_2CuO_2F_{2+\delta}$ and oxidizer-derived Ag.[6,17] The lattice constants were calculated to be $a$ = 3.8880(1) and $c$ = 20.282(1) Å, which are in good agreement with previously reported values.[5] For the data of the sample after annealing with $CuF_2$ (Figure 2(b)), the main XRD peaks were indexed to the orthorhombic *Fmmm* space group (No. 69) with $a$ = 5.525 (1), $b$ = 5.480 (1), and $c$ = 20.250(1) Å. This structural transformation is evidenced by the appearance of the (200) and (020) peaks due to the splitting of the (110) peak of the as-synthesized sample. As displayed in **Figure 2(d)**, the raw material $CuF_2$, which was located beside the sample, partially transformed into CuO through the annealing, which implies that O atoms were extracted from the as-synthesized sample. This annealing method, which is the so-called topochemical fluorination, enables the F and O contents of the samples to be exchanged without affecting the cation framework.[18] Thus, we consider that the fluorination as well as the removal of O simultaneously occurred in this system. Unlike the as-synthesized samples with the *T*-phase, the $CuF_2$-annealed orthorhombic structure was stable for at least two months. Compared with those of the as-synthesized sample (Figure 2(a)), the ($h$00) and (00$l$) peaks of the sample after annealing without $CuF_2$ shifted to lower- and wider-angle sides, respectively, even though its XRD pattern (Figure 2(c)) was also characterized to be of the tetragonal structure. Consequently, the $a$- and $c$-axis lengths were elongated and contracted ($a$ = 3.952 and $c$ = 19.72 Å), respectively. Note that such a change in the lattice constants of the as-synthesized sample was observed even without annealing, i.e., just by keeping it at ambient pressure for an extended period of time (see Figure S2 in the Supporting Information). This suggests that the double-layered $T$-type structure is a metastable phase of $Sr_2SrCu_2O_{4+y}F_{2-y}$. For convenience, hereafter, we will distinguish the samples in Figure 2(b),(c) as the fluorinated sample and the annealed sample, respectively.

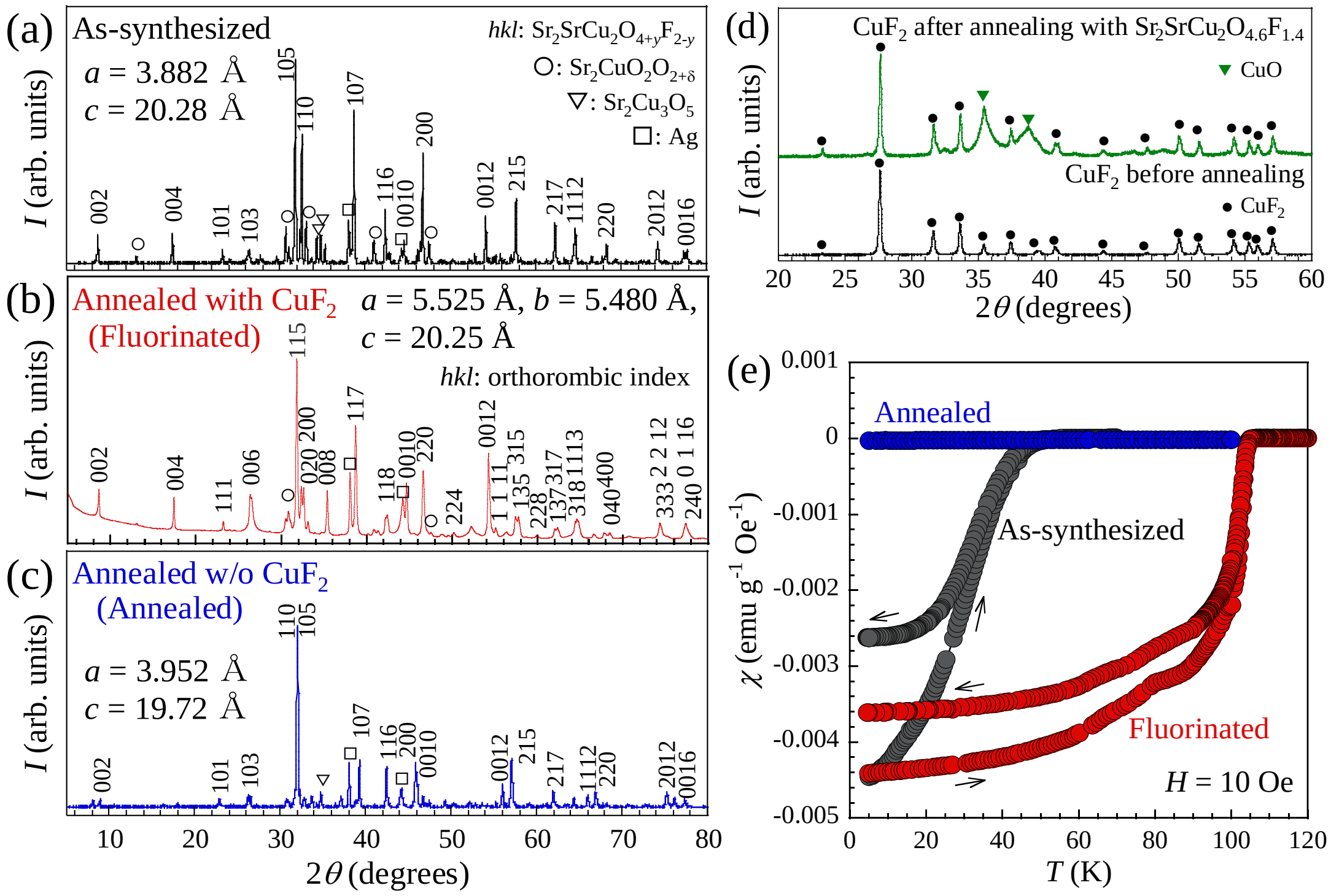

**Figure 2**: Powder X-ray diffraction (XRD) patterns of the (a) as-synthesized sample, (b) $CuF_2$-annealed sample, and (c) sample annealed without $CuF_2$ of $Sr_2SrCu_2O_{4.6}F_{1.4}$. (d) Comparison of the XRD patterns of $CuF_2$ before and after annealing. (e) Experimental $\chi$–$T$ curves for the samples shown in (a)–(c). The arrow toward the higher (lower) $T$ side represents the ZFC (FC) data.

**Figure 2(e)** illustrates the $T$-dependent $\chi$ of each of the samples in Figure 2(a)–(c). With a decrease in $T$, the data of the as-synthesized and fluorinated samples exhibited a rapid decrease toward the lowest temperature. The magnitude of the diamagnetic signal is of the order of ~$4.5\times10^{-3}$ emu/g (approximately 30% of the shielding volume fraction) in both cases, indicating the occurrence of bulk superconductivity. The observed $T_c$ values for the as-synthesized and fluorinated samples are approximately 50 K and 107 K, respectively, the results of which are consistent with previously reported values.[5] Because the structure and superconducting nature of the fluorinated sample is stable in air and ambient pressure, we performed the resistivity measurements of the powdered samples by pressing them under HP (~3.4 GPa). The results indicated that the resistivity exhibits an anomaly originating from the superconducting transition, although its temperature dependence exhibited a nonmetallic behavior (see Figure S3 in the Supporting Information). Interestingly, as shown in Figure 2(e), we found that the annealed sample exhibits no trace of superconductivity above 5 K. By expanding the *y*-axis, a weak diamagnetic signal (approximately $10^{-5}$ emu/g in magnitude) can be observed below ~60 K, which is probably due to the residual *T*-phase in the sample. We assume that the nonsuperconductivity is due to the undoped $Cu^{2+}$ state. This is because the annealed sample, which has a long *a*-axis length of over 3.9 Å, may possess a double-layered structure with a fluorite-type blocking layer (the so-called *T*´-phase), containing no apical oxygen and/or fluorine atoms.

### *3.3. Structure Refinements*

#### *3.3.1 Annealing with $CuF_2$*

The topochemical fluorination of the as-synthesized sample, i.e., low-temperature annealing with $CuF_2$, lowered its structural symmetry from tetragonal to orthorhombic, which led to the removal of excess O and the possible insertion of F. In the single-layered $Sr_2CuO_2F_2$ case, Al-Mamouri et al. reported[6] that the extra F atom was inserted into the interstitial position in the blocking layer by annealing in a $F_2/N_2$ gas atmosphere, resulting in a change to the orthorhombic *Fmmm* structure and an improvement in $T_c$. Here, we performed a structural analysis of the fluorinated $Sr_2SrCu_2O_{4+y}F_{2-y}$ assuming that a similar structural transformation occurred in the present material. As shown in **Figure 3(a)**, we first created a structural model with the orthorhombic *Fmmm* space group, where the extra F atoms were selectively inserted into the interstitial position (named the F2 site in the right panel of Figure 3(a)) surrounded by Sr in the blocking layer. Second, the structural stability of this model was examined using ab initio MD simulations at $T$ = 300 K. The simulation data indicated that the structure was dynamically stable in the duration of the simulation (i.e., 8 ps). Using the constructed model, we performed the Rietveld fitting of the XRD pattern of the fluorinated sample. The best-fit result is shown in **Figure 3(b)**. In the present analysis, the site occupancy (Occ.) of F1, corresponding to the apical fluorine, was fixed at 0.25 to fill each apical site with one F atom. For the F2 site, we performed structural refinements in cases where Occ. = 0, 0.125, and 0.25. The best fit with a weighted-profile reliability factor ($R_{wp}$) of 11.52% and an expected reliability factor ($R_e$) of 6.42% was obtained when the Occ. at the F2 site set to 0.25. The most plausible chemical composition derived from the analysis is $Sr_2SrCu_2O_4F_{2.5}$. The values of the final refinement parameters are listed in **Table I**. The lattice constants are $a$ = 5.5203(1), $b$ = 5.4752(1), and $c$ = 20.2605(4) Å. The $d_{Cu\text{-}Cu}$ distance was calculated to be 3.8874 Å, which is longer than the value of 3.8815 Å (= *a*-axis length) for the as-synthesized sample. By contrast, the *c*-axis was observed to be shorter than that of the as-synthesized sample ($c$ = 20.2843 Å). Consequently, the cell volume increased by 1.2 Å$^3$ from $2V_{tetra}$ = 611.2 Å$^3$ to $V_{ortho}$ = 612.4 Å$^3$, which is possibly because of the F insertion.

The critical structural difference between the fluorinated and as-synthesized samples is that, in the former, the F atom was inserted into the interstitial 8*f* site through fluorination. When this site was excluded from the refined structure, the blocking layer could be regarded as *T*-type. Conversely, when the 32*p* site, corresponding to the apical fluorine, was eliminated, and the 8*f* site was filled, the blocking layer could be regarded as a *T*´-type structure. Therefore, it can be understood that the blocking layer of the fluorinated sample has a structure in which an F atom with Occ. = 0.25 was intercalated into the *T*-type blocking layer.

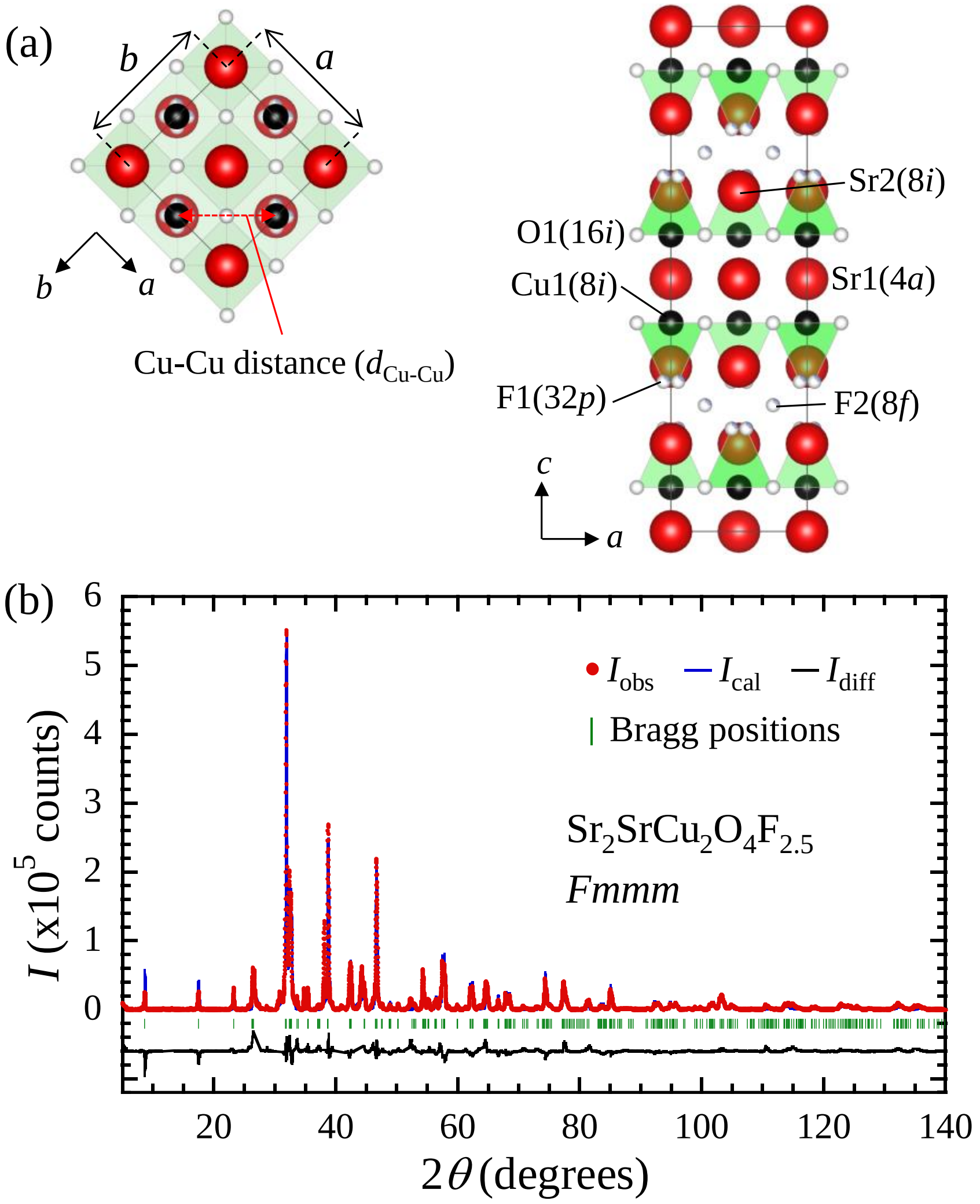

**Figure 3**: (a) Crystal structures of the fluorinated sample possessing the orthorhombic structure with the formula of $Sr_2SrCu_2O_4F_{2.5}$ when viewed along (left) the *c*-axis and (right) the *b*-axis. For visualization, the program VESTA[19] was used. (b) Rietveld refined data of the fluorinated $Sr_2SrCu_2O_{4.4}F_{1.6}$ ($I_{obs}$: observed, $I_{cal}$: calculated). The reliability factor for the fit was $R_{wp}$ = 11.52%.

**Table I**: Atomic coordination data for the fluorinated $Sr_2SrCu_2O_{4.4}F_{4.6}$ measured at $T$ ~ 293 K[a]

| space group: *Fmmm* | | | | | |
|---|---|---|---|---|---|
| $a$ = 5.5203(1) Å, $b$ =5.4752(1) Å, $c$ = 20.2605(4) Å, $V_{orth}$ = 612.4 Å$^3$ | | | | | |
| Atom | WP | $x$ | $y$ | $z$ | Occ. |
| Sr1 | 4*a* | 0 | 0 | 0 | 1 |
| Sr2 | 8*i* | 0 | 0 | 0.1733(1) | 1 |
| Cu1 | 8*i* | 0 | 0 | 0.4128(1) | 1 |
| O1 | 16*j* | 0.25 | 0.25 | 0.4128(1) | 1 |
| F1 | 32*p* | 0.0555 | 0.0532 | 0.2966 | 0.25 |
| F2 | 8*f* | 0.25 | 0.25 | 0.25 | 0.25 |

[a] The atomic displacement parameter, $U_{iso}$, was set to the same value for all atoms and converged to 0.025(1) Å$^2$. The occupancies were fixed for all crystallographic sites. The Sr2–F1 distance was treated as a rigid body and fixed at the DFT-optimized value. Selected interatomic distances: $d_{Sr1-O1}$ = 2.627(2) Å, $d_{Sr2-O1}$ = 2.611(2) Å, $d_{Sr2-F1}$ = 2.535 Å, and $d_{Cu1-F1}$ = 2.391(2) Å.

### *3.3.2. Annealing without $CuF_2$*

When the as-synthesized sample was annealed without $CuF_2$, the crystal structure exhibited anisotropic lattice deformation while maintaining tetragonal symmetry. In particular, the $a$-axis length significantly increased to over 3.9 Å, as mentioned in Section 3.1. Such a long in-plane Cu–Cu distance is characteristic of the $T'$-type blocking layer without F/O atoms at the apical site, as reported for $T'$-$Sr_2CuO_2F_2$ ($a$ ~ 3.97 Å)[20] and $T'$-$La_2CuO_4$ ($a$ ~ 4.02 Å)[21]. Considering these situations, we carried out the Rietveld fitting of the XRD profile of the annealed sample using the $T'$-type structure model with the *I*4/*mmm* space group (see the inset of **Figure 4**). The results are plotted in the main panel of Figure 4. The best fit indicates $a$ = 3.9506(1) and $c$ = 19.6593(8) Å, and the reliability factors converged to $R_{wp}$ = 21.33% and $R_e$ = 10.40%. We also conducted the fitting using the $T$-type structure model; however, the $R_{wp}$ factor was unchanged (possibly because fluorine has relatively fewer electrons than copper and strontium). Therefore, we theoretically examined the phase stability of $Sr_2SrCu_2O_{4+y}F_{2-y}$ assuming double-layered $T$- and $T'$-type structures. **Table II** summarizes the total energy of each system and the Cartesian components of the symmetrized stress tensor. The energy of the $T'$-type structure is lower than that of the $T$-type one by 0.33 eV per primitive cell, which implies that the $T'$-phase of the $Sr_2SrCu_2O_{4+y}F_{2-y}$ system, containing the fluorite-type blocking layer, is rather stable. The optimized fractional coordinates are presented in **Table III**. As shown in the right column of Table II, the $T$-type crystal structure has strongly anisotropic stress: positive and negative pressures along the $a$- and $c$-axis, respectively. Conversely, in the $T'$-structure, the stress is uniformly distributed in each axial direction. These results are consistent with the experimental observations of the *a*-axis elongation and *c*-axis contraction through annealing. Therefore, we conclude that the blocking layer in $Sr_2SrCu_2O_{4+y}F_{2-y}$ changes from *T*- to *T´*-type via annealing to alleviate the anisotropic stress caused by HP synthesis.

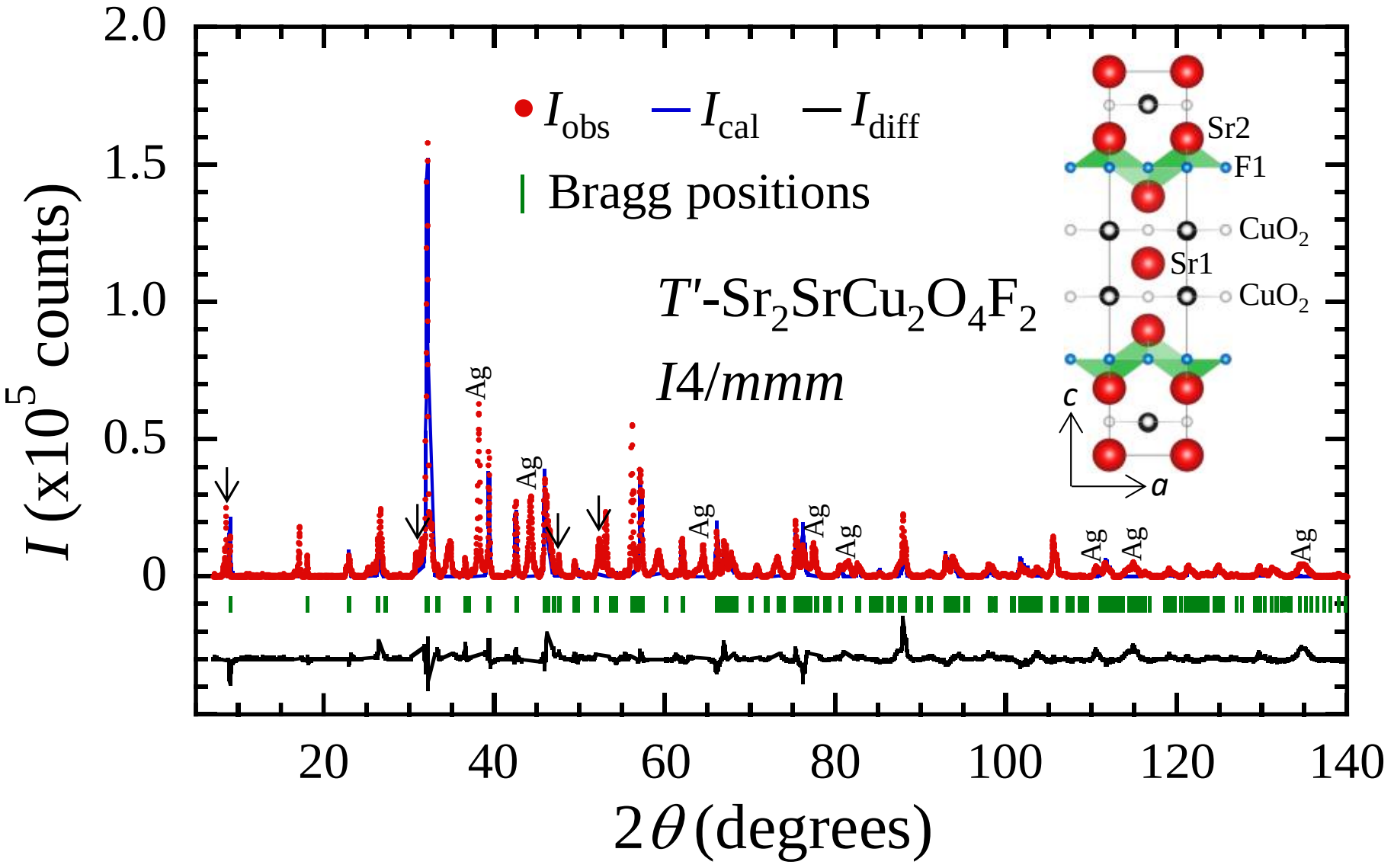

**Figure 4**: Rietveld fit of the sample after annealing without $CuF_2$. The $R_{wp}$ factor was 21.33%. The partial overlap of the peaks of the impurities (arrows) and the target phase is likely to degrade the fitting quality. The crystal structure of the double-layered *T´*-phase of $Sr_2SrCu_2O_4F_2$ is visualized in the inset.

**Table II** Calculated total energy and Cartesian components of the symmetrized stress tensor ($\sigma$) of the *T*- and *T´*-type structures of the stoichiometric $Sr_2SrCu_2O_4F_2$ (*I*4/*mmm*)[a]

| structure types in the blocking layer | total energy (eV/primitive cell) | $\sigma_{xx} = \sigma_{yy}$ (GPa) | $\sigma_{zz}$ (GPa) |
|---|---|---|---|
| *T* (NaCl-type) | –9086.986 | 0.7740 | –5.677 |
| *T´*($CaF_2$-type) | –9087.231 | –2.931 | –2.866 |

[a] $\sigma_{xx}$ ($\sigma_{yy}$) and $\sigma_{zz}$ correspond to the stress in the *a*- and *c*-axis directions, respectively. Due to the symmetry of the tetragonal structure, $\sigma_{xx}$ and $\sigma_{yy}$ are equivalent

**Table III**: Atomic coordination data for $Sr_2SrCu_2O_4F_2$ with the *T´*-type structure, as optimized by the DFT calculation[a]

space group: *I*4/*mmm*

$a$ = 3.9506(1) Å, $c$ = 19.6593(8) Å, $V_{tet}$ = 306.8 Å$^3$

| Atom | WP | *x* | *y* | *z* | *Occ.* |
|---|---|---|---|---|---|
| Sr1 | 2*a* | 0 | 0 | 0 | 1 |
| Sr2 | 4*e* | 0 | 0 | 0.174640 | 1 |
| Cu1 | 4*e* | 0 | 0 | 0.415237 | 1 |
| O1 | 8*g* | 0 | 1/2 | 0.087196 | 1 |
| F1 | 4*d* | 0 | 1/2 | 1/4 | 1 |

[a] The lattice constants were fixed to the experimentally obtained values.

## 4. DISCUSSION

We demonstrated that $Sr_2SrCu_2O_{4+y}F_{2-y}$ exhibits three types of blocking layers: *T*-type, F-intercalated *T*-type, and *T´*-type. **Figure 5** summarizes the structural variation of the $Sr_2SrCu_2O_{4+y}F_{2-y}$ obtained in this study. First, the *T*-type tetragonal structure was stabilized under the HP condition, which yields $T_c$ = 30–60 K.[5] This lower $T_c$ is likely due to the unoptimized hole carrier number. In the present case, the as-synthesized sample is most likely to be in an overdoped state because it was formed by the HP synthesis technique in a strongly oxidizing atmosphere. The lattice constants and $T_c$ changed systematically with an increase in the nominal $y$ of $Sr_2SrCu_2O_{4+y}F_{2-y}$, which indicates an increase in the carrier concentration in the overdoped region (Figure 1). The significantly lower value of the $T_c$ of $Sr_2SrCu_2O_{4+y}F_{2-y}$ than that of $Ba_2CaCu_2O_{4+y}F_{2-y}$ is maybe attributed to the presence of an unexpected chemical disorder in the structure of $Sr_2SrCu_2O_{4+y}F_{2-y}$. As a source of extra hole carriers and disorder, it is possible that the Sr layer between the $CuO_2$ layers accommodates O atoms at random positions. This may encourage further overdoping. The low $T_c$ (~60 K) associated with the partial occupation of O atoms in the Sr(La) layer has been reported in the $La_{2-x}Sr_{1+x}Cu_2O_{6+\delta}$ system.[22,23]

Next, through low-temperature (topochemical) fluorination using $CuF_2$, the structure transformed into an orthorhombic structure, exhibiting a high $T_c$ of 107 K. According to the structural analysis, the fluorination demonstrated that excess apical $O^{2-}$ was removed, and the extra $F^-$ was inserted. Moreover, intercalated $F^-$ occupied the interstitial position in the *T*-type blocking layer. Such an exchange of $O^{2-}$ and $F^-$ indicates a decrease in the formal valence of Cu in $Sr_2SrCu_2O_4(F, O)_2$. Therefore, we conclude that fluorination probably regulates the doping level of the sample from overdoped to nearly optimally doped and, consequently, enhances the $T_c$ in this system. By contrast, the *T*-type blocking layer changes into the *T*´-type when the as-synthesized sample is annealed without $CuF_2$ or exposed to ambient pressure. This route of structural change was first revealed in this study. We also confirmed that the orthorhombic phase could be obtained by annealing the *T*´-phase with $CuF_2$. As reported for *T*´-$Nd_{2-x}Ce_xCuO_4$,[24] an ideal *T*´-phase of $Sr_2SrCu_2O_4F_2$ has the potential to realize *n*-type superconductivity, if the electron carrier is introduced by elemental substitution, e.g., when doping trivalent $La^{3+}$ into the Sr site. To achieve this, further synthesis trials are required.

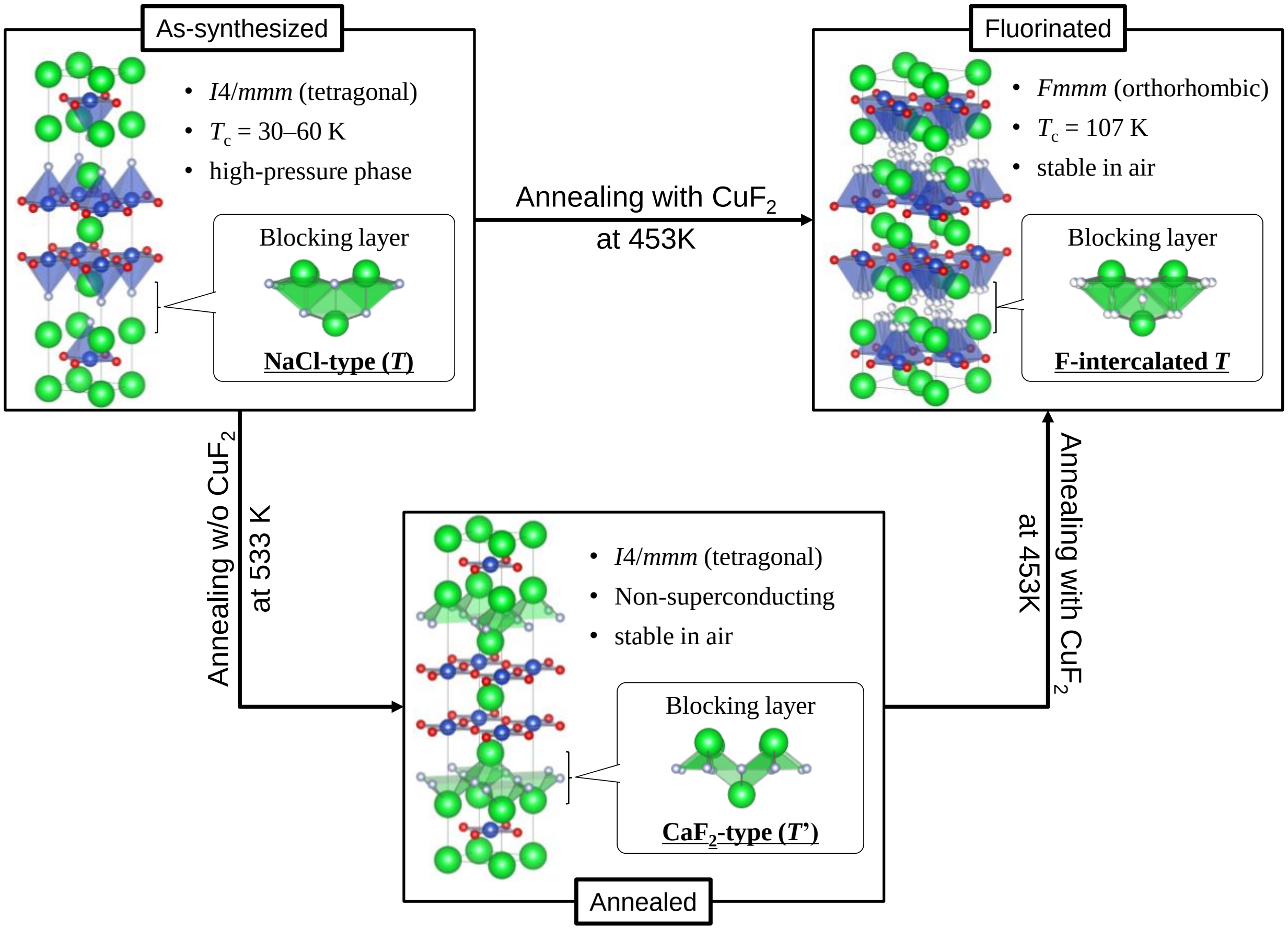

**Figure 5**: Schematic diagram of a structural variation of the $Sr_2SrCu_2O_4F_2$ system depending on the different postannealing methods.

Finally, we discuss the factors that determine the structural selectivity in the blocking layer. In the case of the single-layered oxyfluoride,[25-27] although $Ca_2CuO_2F_2$ has the *T*´-phase, the Sr-replaced $Sr_2CuO_2F_2$ possesses the orthorhombically modulated structure. Conversely, the partially Ba-substituted $Sr_{2-x}Ba_xCuO_2F_{2+\delta}$ ($x > 0.4$) crystallizes in the *T*-phase. According to Pauling's first rule, the coordination number, i.e., the number of anions surrounding a cation, is governed by their ionic radius ratio.[28,29] In the $Sr_2SrCu_2O_{4+y}F_{2-y}$ case, a blocking layer was formed by $Sr^{2+}$ and $F^-$. Moreover, for single- and double-layered cuprates, the $CuO_2$ layers are crystallographically equivalent. Accordingly, the variation in the size of the anions and cations forming the blocking layer directly affects the Cu–O bonding state in the $CuO_2$ layer.

Based on the aforementioned considerations, we plotted the relationship between the in-plane Cu–Cu distance ($d_{Cu–Cu}$) and the radius ratio of the anion/cation in the blocking layer ($r_c/r_a$) in typical oxyfluoride cuprates, as depicted in **Figure 6**. In this case, $r_a$ was fixed as 1.33 Å for $F^-$ (VI)[30]. For comparison, data for the (La, *AE*)$_2$CuO$_4$ system with the simplest *T*-phase is also given with $r_a$ set to 1.40 Å assuming $O^{2-}$(VI). As shown in Figure 6, the preferred configuration of the blocking layer is divided into three regions. When $r_c/r_a$ is larger and $d_{Cu–Cu}$ is shorter, the system tends to settle into the tetragonal *T*-phase (NaCl-type). By contrast, the smaller $r_c/r_a$ and longer $d_{Cu–Cu}$ favor the *T*´-type blocking layer ($CaF_2$-type). Moreover, we found that the orthorhombic regime separated these two regions. For the $Sr_2SrCu_2O_{4+y}F_{2-y}$ system, although the as-synthesized material crystallized in the *T*-phase, the point of ($r_c/r_a$, $d_{Cu–Cu}$) exists in the orthorhombic region. Thus, to overcome the lattice mismatch between the $CuO_2$ and blocking layers, the *T*-phase would change to a more stable *T*´-phase. Comparing the $d_{Cu–Cu}$ lengths of the samples exposed to air and annealed without $CuF_2$, we believe that the latter annealing robustly stabilizes the *T*´-type structure. Furthermore, the fluorinated sample is in the orthorhombic region, which ensures the structural stability of the high-$T_c$ phase in $Sr_2SrCu_2O_{4+y}F_{2-y}$.

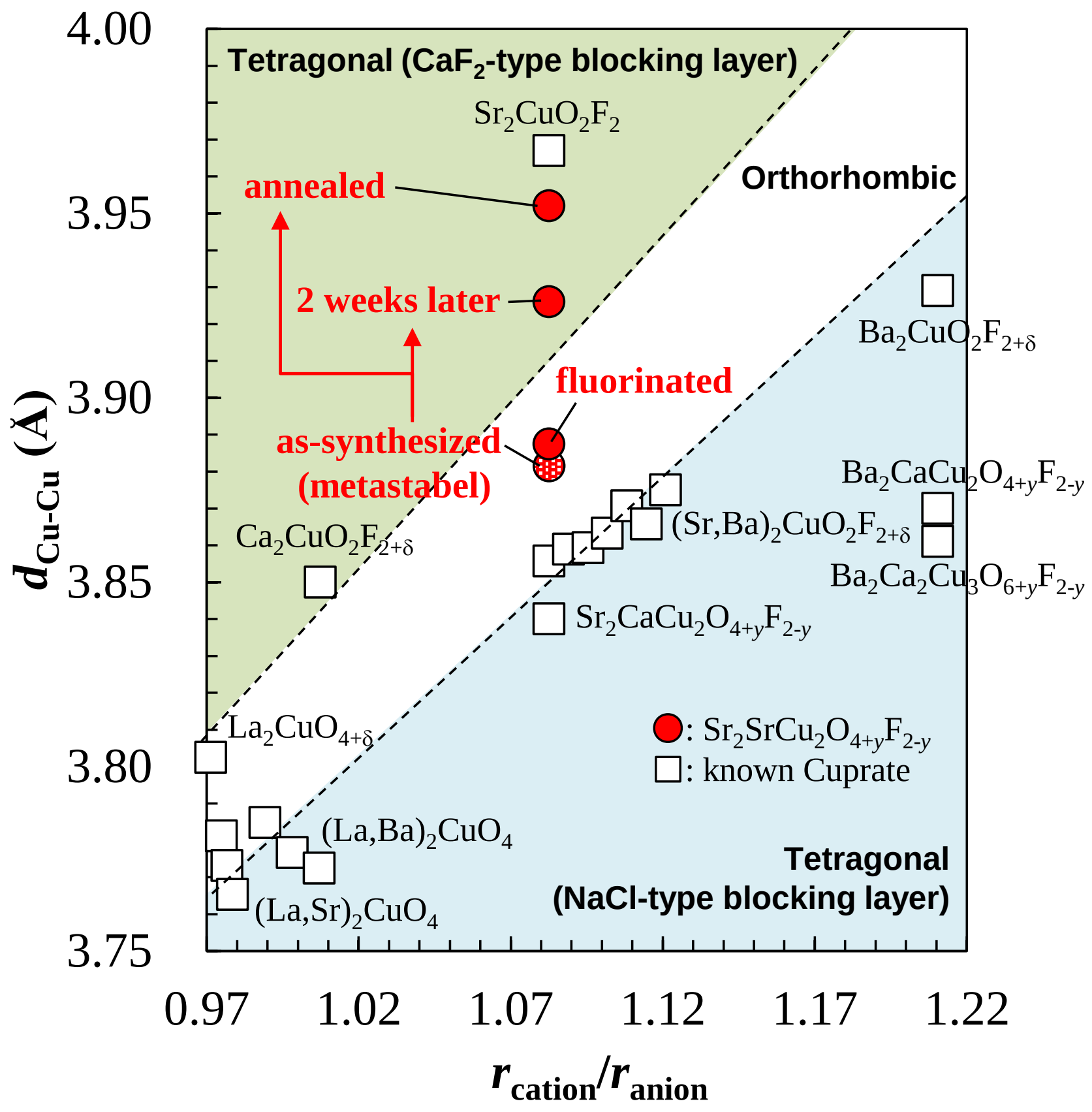

**Figure 6**: Relationship between the in-plane Cu–Cu distance ($d_{Cu–Cu}$) and the cation/anion size ratio ($r_{cation}/r_{anion}$) in the blocking layer for $Sr_2SrCu_2O_{4+y}F_{2-y}$ and other reported cuprates.

5. CONCLUSIONS

We studied various types of crystal structures and the evolution of superconductivity appearing in the Ca-free double-layered oxyfluoride cuprate, $Sr_2SrCu_2O_{4+y}F_{2-y}$. The results suggested that the carrier concentration of the as-synthesized sample (with the tetragonal *T*-phase) can be regulated in the overdoped region by varying the nominal *y* value. The topochemical fluorination resulted in the removal of excess apical $O^{2-}$, insertion of $F^-$, and a remarkable enhancement of $T_c$ to 107 K. We found that the fluorinated phase exhibited orthorhombic *Fmmm* symmetry, whose blocking layer formed a *T*-type structure where the F atom (Occ. = 0.25) was intercalated into the interstitial 8*f* site. Meanwhile, the low-temperature annealing without fluorinating agents altered the blocking layer to the *T´*-phase with a significantly long *a*-axis length of ~3.95 Å. The structural stability of the oxyfluoride cuprates was organized from the viewpoint of lattice matching between the conducting and blocking layers. We found that the relation between the in-plane Cu–Cu distance and the anion/cation combination in the blocking layer is one of the key factors in determining the stability of the crystal structure.

AUTHOR INFORMATION

**Corresponding Author**

*Hiroki Ninomiya, *National Institute of Advanced Industrial Science and Technology (AIST), 1-1-1 Umezono, Tsukuba, Ibaraki 305-8568, Japan*; Email: hiroki.ninomiya@aist.go.jp

**Author Contributions**

‡H.N. and K.K. contributed equally. The manuscript was written through contributions of all authors. All authors have given approval to the final version of the manuscript.

**Funding Sources**

This work was supported by JSPS KAKENHI (grant numbers JP18H01860 and JP19H05823).

**Notes**

The authors declare no competing financial interest.

ACKNOWLEDGMENT

The authors are grateful to Professors Kohji Kishio, Shin-ichi Uchida, and Hideo Aoki at AIST for fruitful discussions.

Supporting Information

# Posttreatment Effects on the Crystal Structure and Superconductivity of Ca-Free Double-Layered Cuprate $Sr_2SrCu_2O_{4+y}F_{2-y}$

Hiroki Ninomiya[1]*‡, Kenji Kawashima[1,2]‡, Hiroshi Fujihisa[1], Shigeyuki Ishida[1], Hiraku Ogino[1], Yoshiyuki Yoshida[1], Hiroshi Eisaki[1], Yoshito Gotoh[1], and Akira Iyo[1]

[1]*National Institute of Advanced Industrial Science and Technology (AIST), 1-1-1 Umezono, Tsukuba, Ibaraki 305-8568, Japan*

[2]*IMRA Japan Co., Ltd., 2-36 Hachiken-cho, Kariya, Aichi 448-8650, Japan.*

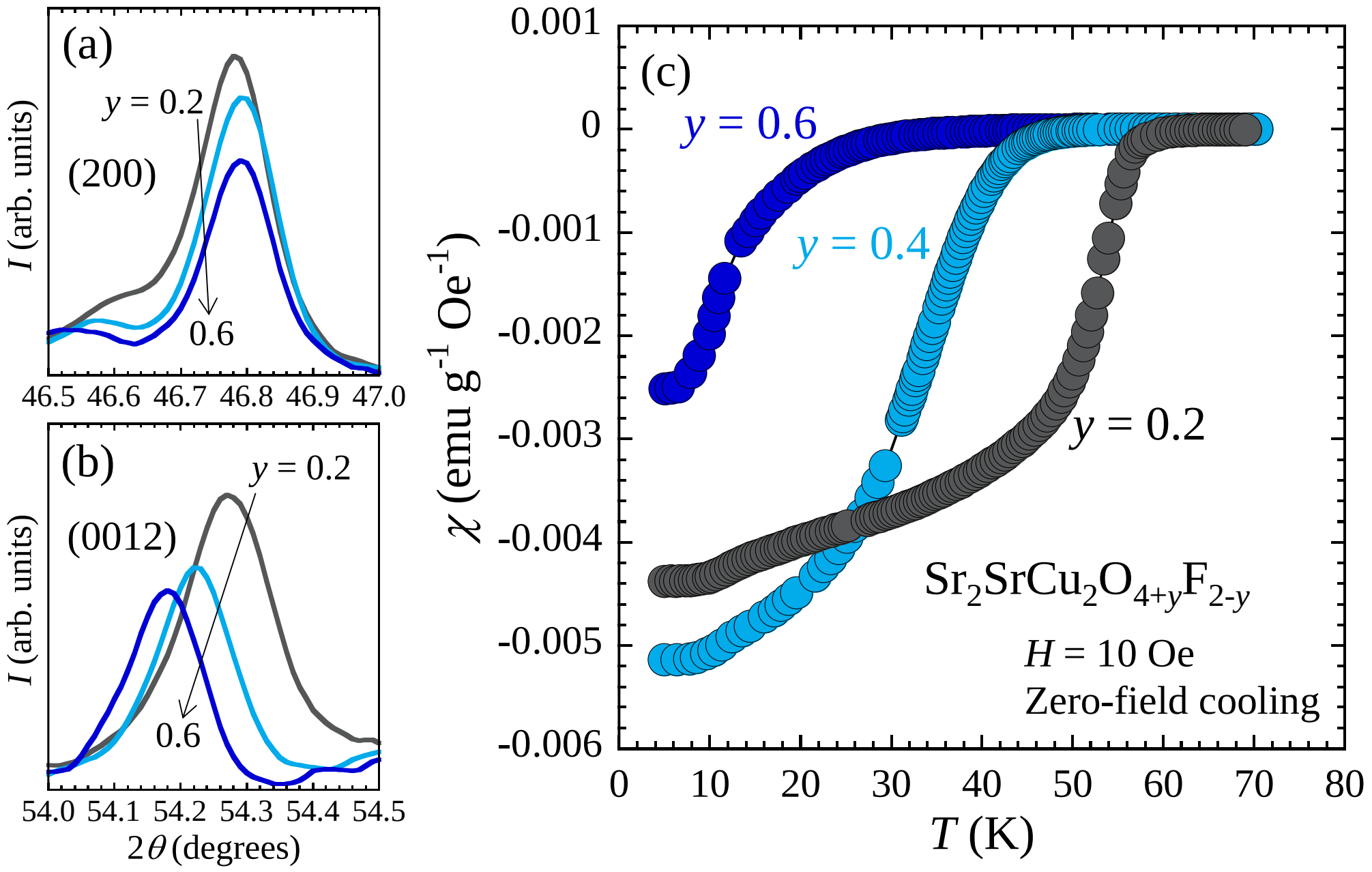

**Figure S1**: Magnified X-ray diffraction patterns at (a) (200) and (b) (0012) peaks of $Sr_2SrCu_2O_{4+y}F_{2-y}$ with $y = 0.2$, 0.4 and 0.6. (c) Temperature dependence of the zero-field-cooled susceptibility of the as-synthesized samples at each $y$.

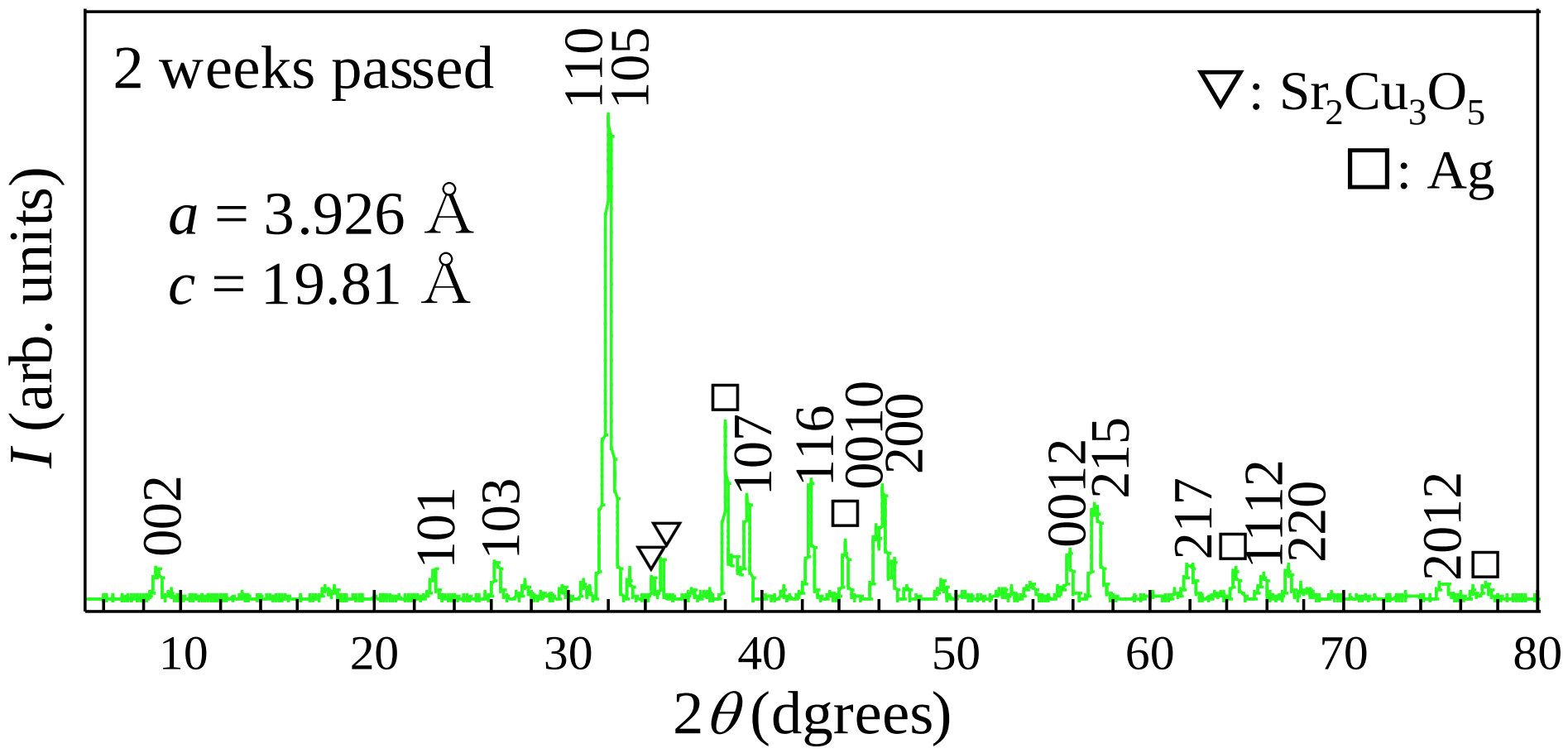

**Figure S2**: Powder X-ray diffraction data of the as-synthesized sample after two weeks at ambient pressure and in air. The main phase is indexed by body-centered tetragonal symmetry, as is the case with the fresh sample shown in Figure 1(a). Compared to the lattice parameters before aging, the *a*- and *c*-axis lengths significantly increased and decreased, respectively.

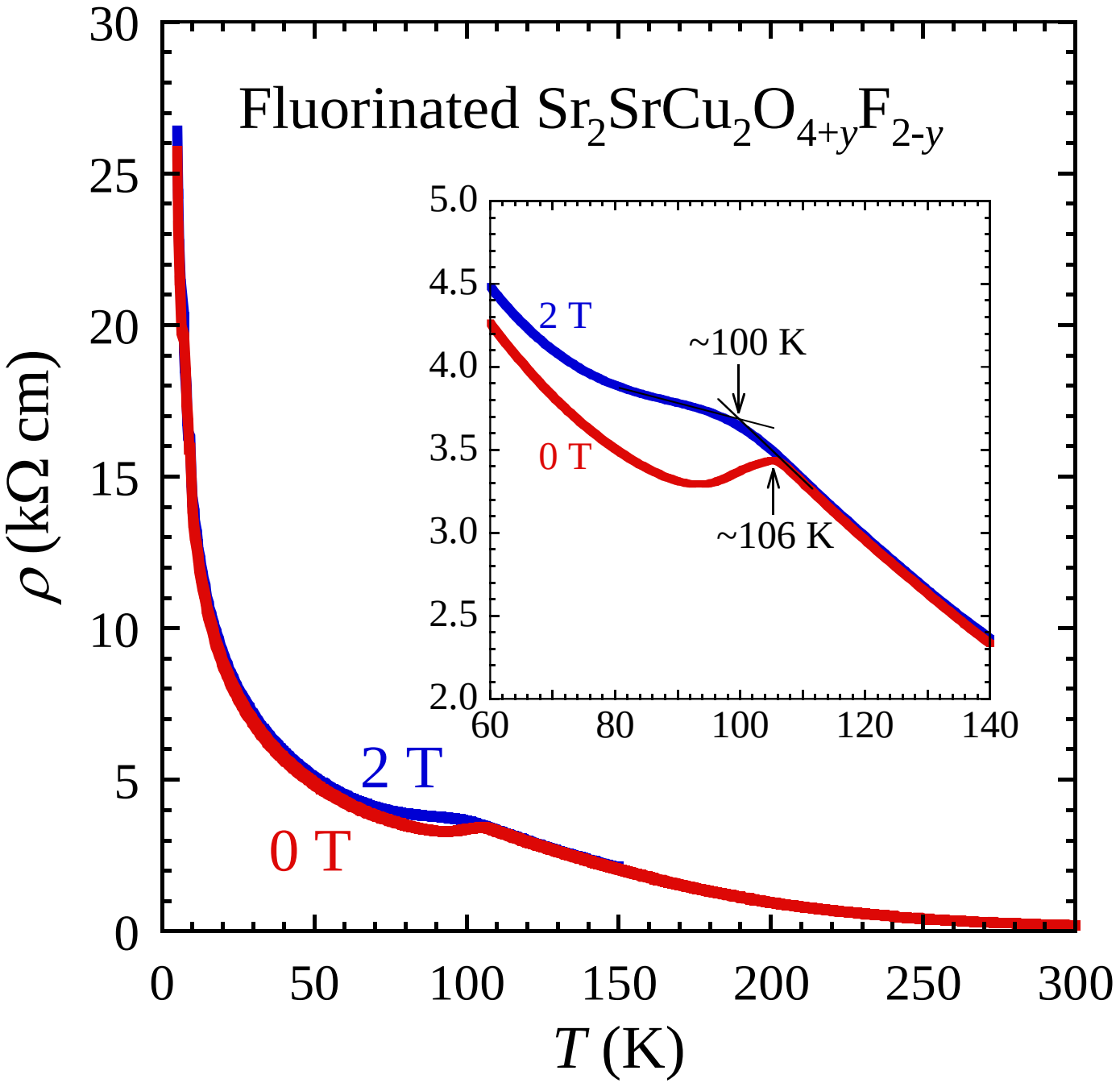

**Figure S3**: Temperature-dependent electrical resistivity of the fluorinated sample measured under the magnetic fields of 0 and 2 T. The inset shows a magnified view at $T_c$.

## Electrical resistivity measurements on the fluorinated sample

Because the structure and superconducting nature of the as-synthesized sample is unstable at the ambient pressure, we performed the resistivity measurements of the fluorinated sample. However, it is difficult to evaluate the transport properties quantitatively, because the low-temperature fluorination was performed using the powdered samples to ensure good chemical homogeneity. In order to solidify the samples after fluorination, they were pressed at 180 °C under 3.4 GPa. The resistivity ($\rho$) data were measured between 5–300 K using the standard four-probe method.

As depicted in **Figure S3**, the resistivity was measured to be 0.2 kΩ cm at 300 K, and exhibited a monotonic increase up to the lowest $T$, revealing a nonmetallic behavior. We also observed that the value reaches its maximum at approximately the $T_c$ (~106 K). Moreover, the peak for the maximum became broader and shifted to a lower $T$ (~ 100 K) under the magnetic field of 2 T. These results indicate that the resistivity anomaly is associated with the transition in superconductivity.